# Main Parameters of the MC and FCC/SppC/LHC/RHIC based Muon-Nucleus Colliders

Burak Dagli[1], Bora Ketenoglu[2], Arif Ozturk[1,*] and Saleh Sultansoy[1,3]

[1] Department of Materials Science and Nanotechnology Engineering, TOBB University of Economics and Technology, Ankara, Turkey
[2] Department of Engineering Physics, Ankara University, Ankara, Turkey
[3] ANAS Institute of Physics, Baku, Azerbaijan

*Corresponding Author: arif.ozturk@etu.edu.tr

**Abstract**

Construction of future Muon Collider (or dedicated $\mu$-ring) tangential to nucleus colliders will give opportunity to realize $\mu A$ collisions at multi-TeV center of mass energies at a luminosity of order of $10^{29}$-$10^{30}$ cm$^{-2}$s$^{-1}$. Obviously, such colliders will essentially enlighten fundamentals of strong interactions from quark to nucleus levels as well as clarify QCD basics. This paper is devoted to estimation of main parameters of the FCC, SppC, HL-LHC, HE-LHC and RHIC based muon-nucleus collider proposals by the software AloHEP.

## 1. Introduction

Lepton-hadron collisions play a crucial role in our understanding of matter's structure (see [1] and references therein). HERA, the first and still unique electron-proton collider, explored structure of protons at $\sqrt{s_{ep}} \approx 0.3\ TeV$ and provided parton distribution functions (PDFs) for adequate interpretation of the LHC and Tevatron data. TeV (or multi-TeV) scale lepton-hadron colliders are crucial for clarifying basics of QCD, which is responsible for 98% of mass of the visible part of our Universe. It should be emphasized that future lepton-hadron colliders will give opportunity to shed light on quark → hadron → nucleus transitions as well. Besides, contruction of TeV energy lepton-hadron colliders is mandatory to provide PDFs for adequate interpretation of forthcoming data from HL/HE-LHC [2,3] and FCC/SppC [4,5].

Today, linac-ring type *ep* colliders are considered as sole realistic way to (multi-) TeV scale in lepton-hadron collisions (see review [6] and references therein) and LHeC [7] is the most promising candidate. However, situation may change in the coming years: $\mu p$ and $\mu A$ colliders can come forward depending on progress in muon beam production and cooling issues. We have reviewed muon-proton collider proposals in [1].

This paper is devoted to estimation of main parameters of the FCC, SppC, HL-LHC, HE-LHC and RHIC based $\mu A$ collider proposals. In section 2, we present different muon-nucleus collider proposals in chronological order. Let us note that FCC based $\mu$-Pb colliders are considered in [8], but mistaken parameters are used for the lead beam; SppC based $\mu$-Pb colliders are considered in [9] for a single energy choice for muon beam; the RHIC based $\mu$-Au colliders have been proposed in [10,11] without luminosity estimations; HL-LHC and HE-LHC based $\mu$-Pb colliders are considered in this paper for the first time. Main parameters of all these muon-nucleus colliders are computed by the software AloHEP [12–15] that is developed for quick estimation of luminosities, tune shifts and disruption for different types of colliders. Finally, in Section 3 we present our conclusions and recommendations.

## 2. Muon-nucleus colliders

In this section, we present main parameters of different muon-nucleus colliders.

### 2.1. FCC based *μ-Pb* Colliders

FCC based energy-frontier lepton-hadron colliders have been proposed in [8], where main parameters of *ep*, *μp*, *e-Pb* and *μ-Pb* colliders are estimated. As mentioned in Introduction, mistaken values for lead beam have been used. Main parameters of the FCC based *μ-p* colliders are cross-checked by using AloHEP in [1]. In this subsection, parameters of FCC based *μ-Pb* colliders are computed by using AloHEP software (corrected values of the FCC based *e-Pb* option will be presented in a separate paper).

Main parameters of muon beam are summarized in Table 1 [16]. These parameters will be used in subsections 2.2, 2.3 and 2.4 as well.

**Table 1.** Muon Collider parameters

| Options | I | II | III |
|---|---|---|---|
| Beam Energy [TeV] | 0.75 | 1.5 | 3.0 |
| Circumference [km] | 2.5 | 4.5 | 6.0 |
| Particle per Bunch [$10^{12}$] | 2 | 2 | 2 |
| Repetition rate [Hz] | 15 | 12 | 6 |
| Normalized Emittance [μm] | 25 | 25 | 25 |
| $\beta^*$ function at IP [cm] | 1 | 0.5 | 0.25 |
| IP beam size [μm] | 5.9 | 3.0 | 1.48 |
| Bunch length, $\sigma_z$ [cm] | 1 | 0.5 | 0.2 |
| Bunches per beam | 1 | 1 | 1 |

Nominal parameters of the FCC *Pb-Pb* collider (Table 2.10 in [4]) are presented in the second column of Table 2. Parameters of lead beam upgraded for ERL60 and FCC based *e-Pb* collider (Table 2.13 in [4]) are given in the last column of Table 2.

**Table 2.** FCC lead beam parameters

| Options | Pb-Pb | e-Pb |
|---|---|---|
| Beam Energy [TeV] | 4100 | 4100 |
| Circumference [km] | 100 | 100 |
| Particle per Bunch [$10^8$] | 2 | 1.8 |
| Revolution freq. [Hz] | 2998 | 2998 |
| Normalized Emittance [μm] | 1.5 | 0.9 |
| $\beta^*$ function at IP [cm] | 30 | 15 |
| Bunches per beam | 5400 | 2072 |
| Bunch length [cm] | 8 | 8 |

Implementation of these parameters into the software AloHEP leads to results presented in Table 3 for nominal/ERL60 upgraded lead beam parameters.

**Table 3.** FCC based *μ-Pb* collider parameters with nominal/ERL60 upgraded lead beam

| Options | I | II | III |
|---|---|---|---|
| $\sqrt{s}$ [TeV] | 111 | 157 | 222 |
| IP beam size [μm] | 5.94/5.94 | 4.62/2.97 | 4.62/2.53 |
| Tune shift $\xi_{Pb}$ | 5.28/8.81 | | |
| Tune shift $\xi_\mu$ [$10^{-4}$] | 7.11/6.40 | | |
| Peak L [$10^{30}$cm$^{-2}$s$^{-1}$] | 2.52/2.28 | 3.72/8.09 | 2.78/8.35 |
| L with decay [$10^{30}$cm$^{-2}$s$^{-1}$] | 1.60/1.44 | 2.34/5.12 | 1.76/5.28 |

It is seen that beam-beam parameter for lead is very high. One way to reduce it to acceptable value 0.01 is 528/881 times decreasing of number of muons per bunch which will cause corresponding reducing in luminosity. This reducing may be partially compensated by increasing the number of Pb per bunch. Increasing of this number by factor 200 results in value 0.14/0.13 for muon beam tune shift which seems acceptable for μ-Pb colliders. In summary, these modifications lead to approximately 2.5/4.4 times reduction of luminosity values (see Table 4).

**Table 4.** FCC based *μ-Pb* collider parameters with upgraded numbers of μ and Pb per bunch

| Options | I | II | III |
|---|---|---|---|
| $\sqrt{s}$ [TeV] | 111 | 157 | 222 |
| $N_\mu$ [$10^9$] | 3.79/2.28 | | |
| $N_{Pb}$ [$10^{10}$] | 4/3.6 | | |
| IP beam size [μm] | 5.94/5.94 | 4.62/2.97 | 4.62/2.53 |
| Tune shift $\xi_{Pb}$ | 0.01 | | |
| Tune shift $\xi_\mu$ | 0.14/0.13 | | |
| Peak L [$10^{29}$ cm$^{-2}$s$^{-1}$] | 9.58/5.19 | 14.1/18.5 | 10.6/19.0 |
| L with decay [$10^{29}$ cm$^{-2}$s$^{-1}$] | 6.06/3.28 | 8.89/11.7 | 6.67/12.0 |

### 2.2. SppC based *μ-Pb* Colliders

SppC based *μ-Pb* collider has been considered for a single muon energy option ($E_\mu$ = 1.5 TeV) in [9]. Here we evaluate all three options given in Table 1. For lead beam parameters we use values presented in Table 5 (see Table A7.2 in [5]).

**Table 5.** SppC lead beam parameters

| Beam Energy [TeV] | 3075 |
|---|---|
| Circumference [km] | 100 |
| Particle per Bunch [$10^9$] | 1.8 |
| Revolution rate [Hz] | 3000 |
| Normalized Emittance [μm] | 0.22 |
| $\beta^*$ function at IP [cm] | 75 |
| IP beam size [μm] | 3.25 |
| Bunches per beam | 10080 |

Implementing Tables 1 and 5 into the software AloHEP, we have obtained parameters of *μ-Pb* collisions summarized in Table 6. Let us mention that peak luminosity value for option II is 3 times less than the value given in [9]. The reason for this is that collision frequency is taken equal to the revolution frequency of muon collider, namely 66 kHz. Instead, the collision frequency value of 21.6 kHz has to be taken into account, assuming the muon beam repetition rate 12 Hz and approximately 1800 revolutions per pulse train.

**Table 6.** SppC based *μ-Pb* collider parameters

| Options | I | II | III |
|---|---|---|---|
| √s [TeV] | 96 | 136 | 192 |
| IP beam size [μm] | 5.94 | 3.23 | 3.23 |
| Tune shift $\xi_{Pb}$ | 36 | | |
| Tune shift $\xi_\mu$ [$10^{-3}$] | 6.4 | | |
| Peak L [$10^{31}$cm$^{-2}$s$^{-1}$] | 2.28 | 6.83 | 5.12 |
| L with decay [$10^{31}$cm$^{-2}$s$^{-1}$] | 1.44 | 4.32 | 3.24 |

It is seen that beam-beam parameter for lead is extremely high. It can be reduced to an acceptable value of 0.01 by decreasing number of muons 3600 times per bunch, which will result in a corresponding reduction in luminosity as a matter of course. This reduction may partially be compensated by increasing the number of Pb per bunch. Increase of this number by a factor of 16 results in tune shift value of 0.1. In summary, these modifications lead to 225 times reduction of luminosity values (see Table 7).

**Table 7.** SppC based *μ-Pb* collider parameters with upgraded numbers of *μ* and *Pb* per bunch

| Options | I | II | III |
|---|---|---|---|
| √s [TeV] | 96 | 136 | 192 |
| $N_\mu$ [$10^8$] | 5.56 | | |
| $N_{Pb}$ [$10^{10}$] | 2.88 | | |
| IP beam size [μm] | 5.94 | 3.23 | 3.23 |
| Tune shift $\xi_{Pb}$ | 0.01 | | |
| Tune shift $\xi_\mu$ | 0.1 | | |
| Peak L [$10^{29}$ cm$^{-2}$s$^{-1}$] | 1.01 | 3.04 | 2.28 |
| L with decay [$10^{29}$ cm$^{-2}$s$^{-1}$] | 0.64 | 1.92 | 1.44 |

### 2.3. HL-LHC based *μ-Pb* Colliders

HL-LHC based *μp* colliders were proposed in [17]. In this subsection we extend this proposal to *μA* option.

Parameters of lead beam upgraded for ERL60 and HL-LHC based *e-Pb* collider (Table 2.13 in [4]) are given in Table 8.

**Table 8.** HL-LHC lead beam parameters

| Beam Energy [TeV] | 574 |
|---|---|
| Circumference [km] | 26.7 |
| Particle per Bunch [$10^8$] | 1.8 |
| Revolution freq. [Hz] | 11245 |
| Normalized Emittance [μm] | 1.5 |
| $\beta^*$ function at IP [cm] | 7 |
| Bunches per beam | 1200 |
| Bunch length [cm] | 8 |

Implementing Tables 1 and 8 into the software AloHEP, we have obtained parameters of *μ-Pb* collisions summarized in Table 9.

**Table 9.** HL-LHC based *μ-Pb* collider parameters

| Options | I | II | III |
|---|---|---|---|
| √s [TeV] | 41.5 | 58.7 | 82.9 |
| IP beam size [μm] | 5.96 | | |
| Tune shift $\xi_{Pb}$ | 5.28 | | |
| Tune shift $\xi_\mu$ [$10^{-4}$] | 6.40 | | |
| Peak L [$10^{30}$cm$^{-2}$s$^{-1}$] | 2.25 | 2.00 | 1.50 |
| L with decay [$10^{30}$cm$^{-2}$s$^{-1}$] | 1.42 | 1.27 | 0.95 |

It is seen that beam-beam parameter for lead is very high. One way to reduce it to acceptable value 0.01 is 528 times decreasing of number of muons per bunch which will cause corresponding reducing in luminosity. This reducing may be partially compensated by increasing the number of *Pb* per bunch. Increasing of this number by factor 160 results in value 0.1 for muon beam tune shift. In summary, these modifications lead to approximately 3.3 times reduction of luminosity values (see Table 10).

**Table 10.** HL-LHC based *μ-Pb* collider parameters with upgraded numbers of *μ* and *Pb* per bunch

| Options | I | II | III |
|---|---|---|---|
| √s [TeV] | 41.5 | 58.7 | 82.9 |
| $N_\mu$ [$10^9$] | 3.79 | | |
| $N_{Pb}$ [$10^{10}$] | 2.88 | | |
| IP beam size [μm] | 5.96 | | |
| Tune shift $\xi_{Pb}$ | 0.01 | | |
| Tune shift $\xi_\mu$ | 0.1 | | |
| Peak L [$10^{29}$ cm$^{-2}$s$^{-1}$] | 6.83 | 6.07 | 4.56 |
| L with decay [$10^{29}$ cm$^{-2}$s$^{-1}$] | 4.32 | 3.84 | 2.88 |

### 2.4. HE-LHC based *μ-Pb* Colliders

HE-LHC based *μp* colliders were proposed in [17]. In this subsection we extend this proposal to *μA* option.

Parameters of lead beam upgraded for ERL60 and HE-LHC based *e-Pb* collider (Table 2.13 in [4]) are given in Table 11.

**Table 11.** HE-LHC lead beam parameters

| Beam Energy [TeV] | 1030 |
|---|---|
| Circumference [km] | 26.7 |
| Particle per Bunch [$10^8$] | 1.8 |
| Revolution freq. [Hz] | 11245 |
| Normalized Emittance [μm] | 1 |
| $\beta^*$ function at IP [cm] | 10 |
| Bunches per beam | 1200 |
| Bunch length [cm] | 8 |

Implementing Tables 1 and 11 into the software AloHEP, we have obtained parameters of *μ-Pb* collisions summarized in Table 12.

**Table 12.** HE-LHC based μ-Pb collider parameters

| Options | I | II | III |
|---|---|---|---|
| √s [TeV] | 55.6 | 78.6 | 111 |
| IP beam size [μm] | 5.94 | 4.34 | 4.34 |
| Tune shift $\xi_{Pb}$ | | 7.92 | |
| Tune shift $\xi_\mu$ [$10^{-4}$] | | 6.40 | |
| Peak L [$10^{30}$ cm$^{-2}$s$^{-1}$] | 2.28 | 3.77 | 2.83 |
| L with decay [$10^{30}$ cm$^{-2}$s$^{-1}$] | 1.44 | 2.39 | 1.79 |

It is seen that beam-beam parameter for lead is very high. It can be reduced to an acceptable value of 0.01 by decreasing number of muons 792 times per bunch, which will result in a corresponding reduction in luminosity as a matter of course. This reduction may partially be compensated by increasing the number of *Pb* per bunch. Increase of this number by a factor of 160 results in tune shift value of 0.1 for muon beam which is acceptable for *μ-Pb* colliders. In summary, these modifications lead to 5 times reduction of luminosity values (see Table 13).

**Table 13.** HE-LHC based μ-Pb collider parameters with upgraded numbers of μ and *Pb* per bunch

| Options | I | II | III |
|---|---|---|---|
| √s [TeV] | 55.6 | 78.6 | 111 |
| $N_\mu$ [$10^9$] | | 2.53 | |
| $N_{Pb}$ [$10^{10}$] | | 2.88 | |
| IP beam size [μm] | 5.94 | 4.34 | 4.34 |
| Tune shift $\xi_{Pb}$ | | 0.01 | |
| Tune shift $\xi_\mu$ | | 0.1 | |
| Peak L [$10^{29}$ cm$^{-2}$s$^{-1}$] | 4.61 | 7.64 | 5.73 |
| L with decay [$10^{29}$ cm$^{-2}$s$^{-1}$] | 2.91 | 4.83 | 3.62 |

### 2.5. RHIC based μ-Au Colliders

The RHIC/EIC based muon-nucleus collider has been proposed in [10] assuming reusage of the electron ring for acceleration of muons up to 0.96 TeV. Later on, other muon energy options have been considered in [11]. In References [10,11], parameters of μp colliders are presented, whereas μA parameters are not given. In this subsection, we evaluate main parameters of μ-Au collisions for $E_\mu$ = 0.96 TeV option. Parameters of muon beam are presented in Table 14 (see Table 2 of [10]).

**Table 14.** Parameters of muon beam

| Beam Energy [TeV] | 0.96 |
|---|---|
| Circumference [km] | 3.834 |
| Particle per Bunch [$10^{11}$] | 20 |
| Repetition rate [Hz] | 15 |
| Normalized Emittance [μm] | 25 |
| $\beta^*$ function at IP [cm] | 1 |
| IP beam size [μm] | 5.2 |
| Bunches per beam | 1 |

Regarding gold beam parameters, values from Table 32.5 in [18] are summarized in Table 15.

**Table 15.** Parameters of Gold beam

| Beam Energy [TeV] | 19.7 |
|---|---|
| Circumference [km] | 3.834 |
| Particle per Bunch [$10^9$] | 2 |
| Revolution rate [Hz] | 78250 |
| Normalized Emittance [μm] | 2.23 |
| $\beta^*$ function at IP [cm] | 70 |
| Bunches per beam | 111 |

Implementing Tables 14 and 15 into the software AloHEP, we have obtained parameters of μ-Au collisions presented in Table 16.

**Table 16.** RHIC based μ-Au collider parameters

| √s [TeV] | 8.7 |
|---|---|
| IP beam size [μm] | 121 |
| Tune shift $\xi_{Au}$ | 3.47 |
| Tune shift $\xi_\mu$ [$10^{-3}$] | 6.85 |
| Peak L [$10^{29}$ cm$^{-2}$s$^{-1}$] | 1.08 |
| L with decay [$10^{29}$ cm$^{-2}$s$^{-1}$] | 0.45 |

It is seen that beam-beam parameter for gold is very high. It can be reduced to an acceptable value of 0.01 by decreasing number of muons 347 times per bunch, which will result in a corresponding reduction in luminosity as a matter of course. This reduction may partially be compensated by increasing the number of *Au* per bunch. Increase of this number by a factor of 14.6 results in tune shift value of 0.1 for muon beam which is acceptable for *μ-Au* colliders. In summary, these modifications lead to 24 times reduction of luminosity values (see Table 17).

**Table 17.** RHIC based μ-Au collider parameters with upgraded numbers of μ and *Au* per bunch

| √s [TeV] | 8.69 |
|---|---|
| $N_\mu$ [$10^9$] | 5.76 |
| $N_{Au}$ [$10^{10}$] | 2.92 |
| IP beam size [μm] | 121 |
| Tune shift $\xi_{Au}$ | 0.01 |
| Tune shift $\xi_\mu$ | 0.1 |
| Peak L [$10^{27}$ cm$^{-2}$s$^{-1}$] | 4.53 |
| L with decay [$10^{27}$ cm$^{-2}$s$^{-1}$] | 1.88 |

### 3. Conclusion

Construction of future muon collider (or dedicated μ-ring) tangential to existing and proposed hadron colliders will give opportunity to realize μA colliders with multi-TeV center-of-mass energies at a luminosity of order of $10^{29}$-$10^{30}$ cm$^{-2}$s$^{-1}$ for FCC, SppC, HL-LHC, HE-LHC based and $10^{27}$ cm$^{-2}$s$^{-1}$ for RHIC based muon-nucleus colliders. Obviously, such colliders will essentially enlarge the physics search potential of corresponding nucleus colliders.

As seen in Section 2, needed number of muons per bunch for $\mu A$ colliders are 2-3 orders less than that of muon colliders. In this respect, construction of $\mu A$ colliders seems easier than MC itself. Therefore, the following strategy can be foreseen: construction of the muon ring tangential to the hadron colliders to investigate $\mu p$ and $\mu A$ collisions as a first step and then realization of Muon Collider in the same ring.

Finally, let us remind that the EMC effect was discovered in muon-nucleus scattering experiments. The proposed $\mu A$ colliders will provide opportunity to study this phenomenon in a much larger kinematic region. Overall, these colliders will pave the way for understanding nature of strong interactions at the nuclear level.

Scientific potential of the RHIC based muon-nucleus collider has been examined in [10,11]. Similar studies should be done for FCC, SppC, HL-LHC, HE-LHC based $\mu A$ colliders as well.

**Acknowledgements**

The authors are grateful to Umit Kaya and Bilgehan Baris Oner for their contributions in early stages of the AloHEP software.

**References**

[1] Dagli B, Ketenoglu B and Sultansoy S 2022 Review of Muon-Proton Collider Proposals: Main Parameters *ArXiv Prepr. ArXiv220600037*

[2] Apollinari G, Béjar Alonso I, Brüning O, Fessia P, Lamont M, Rossi L and Tavian L 2017 *High-Luminosity Large Hadron Collider (HL-LHC). Technical Design Report V. 0.1* (Fermi National Accelerator Lab.(FNAL), Batavia, IL (United States))

[3] Abelleira J L, Amorim D, Antipov S A, Apyan A, Arsenyev S, Barranco J, Benedikt M, Bruce R, Burkart F and Cai Y 2018 High-energy LHC design *Journal of Physics: Conference Series* vol 1067 (IOP Publishing) p 022009

[4] Benedikt M, Mertens V, Cerutti F, Riegler W, Otto T, Tommasini D, Tavian L J, Gutleber J, Zimmermann F and Mangano M 2018 FCC-hh: The Hadron Collider: future circular collider conceptual design report volume 3 *Eur Phys J Spec Top* **228** 755–1107

[5] Group C S 2018 CEPC conceptual design report: Volume 1-accelerator *ArXiv Prepr. ArXiv180900285*

[6] Akay A N, Karadeniz H and Sultansoy S 2010 Review of linac–ring-type collider proposals *Int. J. Mod. Phys. A* **25** 4589–602

[7] Fernandez J A, Adolphsen C, Akay A N, Aksakal H, Albacete J L, Alekhin S, Allport P, Andreev V, Appleby R B and Arikan E 2012 A large hadron electron collider at CERN report on the physics and design concepts for machine and detector *J. Phys. G Nucl. Part. Phys.* **39** 075001

[8] Acar Y C, Akay A N, Beser S, Canbay A C, Karadeniz H, Kaya U, Oner B B and Sultansoy S 2017 Future circular collider based lepton–hadron and photon–hadron colliders: Luminosity and physics *Nucl. Instrum. Methods Phys. Res. Sect. Accel. Spectrometers Detect. Assoc. Equip.* **871** 47–53

[9] Ketenoglu B 2021 Main parameters of SppC-based "linac-ring eA" and "ring-ring μA" colliders *Can. J. Phys.* **99** 259–62

[10] Acosta D and Li W 2022 A muon–ion collider at BNL: The future QCD frontier and path to a new energy frontier of μ+ μ- colliders *Nucl. Instrum. Methods Phys. Res. Sect. Accel. Spectrometers Detect. Assoc. Equip.* **1027** 166334

[11] Acosta D, Barberis E, Hurley N, Li W, Colin O M, Wood D and Zuo X 2022 The Potential of a TeV-Scale Muon-Ion Collider *ArXiv Prepr. ArXiv220306258*

[12] Kaya U and Oner B B 2017 A Luminosity Optimizer for High Energy Physics (version 1) TOBB ETU High Energy Physics Group http://yef.etu.edu.tr/ALOHEP_eng.html

[13] Dagli B and Oner B B 2020 A Luminosity Optimizer for High Energy Physics (version 2) TOBB ETU High Energy Physics Group http://yef.etu.edu.tr/ALOHEP_eng2.html

[14] Dagli B, Sultansoy S, Ketenoglu B and Oner B B 2021 Beam-beam Simulations for Lepton-hadron Colliders: AloHEP Software *12th Int Part. Acc Conf IPAC'21*

[15] Dagli B and Ozturk A 2022 A Luminosity Optimizer for High Energy Physics (AloHEP) *https://github.com/yefetu/ALOHEP*

[16] Delahaye J P, Diemoz M, Long K, Mansoulié B, Pastrone N, Rivkin L, Schulte D, Skrinsky A and Wulzer A 2019 Muon colliders *ArXiv Prepr. ArXiv190106150*

[17] Kaya U, Ketenoglu B, Sultansoy S and Zimmermann F 2022 Luminosity and physics considerations on HL-LHC– and HE-LHC–based μp colliders *Europhys. Lett.* **138** 24002

[18] Zyla (Particle Data Group) P A *Prog. Theor. Exp. Phys.* 2020, 083C01 (2020) and 2021 update